\documentclass[12pt]{amsart}
\usepackage{enumerate}
\usepackage{enumitem}
\usepackage[colorlinks=true, linkcolor=blue, urlcolor=blue, citecolor=blue, anchorcolor=blue, pdfborder={0 0 0}]{hyperref}
\usepackage{url}
\usepackage{breqn}
\usepackage{graphicx,color,colortbl}
\usepackage[dvipsnames,table,xcdraw]{xcolor}
\usepackage{cite}
\usepackage{amsthm, amsmath, amssymb}
\usepackage{mathtools}
\usepackage[top=45truemm, bottom=45truemm, left=30truemm, right=30truemm]{geometry}
\usepackage{nicefrac}
\usepackage{cancel}
\usepackage{float}
\usepackage{tabularx}
\usepackage{makecell}
\usepackage{array}
\usepackage{ragged2e}
\usepackage{graphicx}
\usepackage{caption}
\usepackage{subcaption}
\usepackage[symbol]{footmisc}
\usepackage{tikz}
\makeatletter
\def\@settitle{\begin{center}%
  \baselineskip14\p@\relax
  \bfseries
  \uppercasenonmath\@title
  \@title
  \ifx\@subtitle\@empty\else
     \\[1ex]\uppercasenonmath\@subtitle
     \footnotesize\mdseries\@subtitle
  \fi
  \end{center}%
}
\def\subtitle#1{\gdef\@subtitle{#1}}
\def\@subtitle{}
\makeatother

\definecolor{bg}{rgb}{0.95,0.95,0.95}
\definecolor{LightGray}{RGB}{242,242,242}

\newcolumntype{P}[1]{>{\RaggedRight\hspace{0pt}}p{#1}}
\newcolumntype{L}[1]{>{\raggedright\let\newline\\\arraybackslash\hspace{0pt}}m{#1}}
\newcolumntype{C}[1]{>{\centering\let\newline\\\arraybackslash\hspace{0pt}}m{#1}}
\newcolumntype{R}[1]{>{\raggedleft\let\newline\\\arraybackslash\hspace{0pt}}m{#1}}

\newtheorem{definition}{Definition}

\theoremstyle{definition}

\title{Which programming languages do hackers use?} 
\subtitle{A survey at the German Chaos Computer Club}

\author[Christian Koch]{Christian Koch}
\address{Christian Koch\\Technische Hochschule Nürnberg Georg Simon Ohm\\\mbox{Kesslerpl. 12}\\Nuremberg\\Germany}
\curraddr{}
\email{christian.koch@th-nuernberg.de}

\author[Katharina Müller]{Katharina Müller}
\address{Katharina Müller\\ Friedrich-Alexander University Erlangen-Nuremberg\\\mbox{Martensstr. 5a}\\Erlangen\\Germany}
\curraddr{}
\email{katharina.mueller@mnet-mail.de}

\author[Eldar Sultanow]{\href{https://orcid.org/0000-0001-5257-2236}{\includegraphics[scale=0.06]{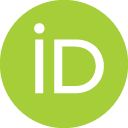}}\hspace{1mm}Eldar Sultanow}
\address{Eldar Sultanow\\Chair of Business Informatics, Processes and Systems\\Potsdam University\\Karl-Marx Straße 67\\14482 Potsdam\\Germany}
\curraddr{}
\email{eldar.sultanow@wi.uni-potsdam.de}


\begin{document}

\begingroup
\let\MakeUppercase\relax
\clearpage\maketitle
\thispagestyle{empty}
\endgroup

\begin{abstract}
There are numerous articles about the programming languages most commonly used by hackers. Among them, however, there are hardly any scientific studies. One reason might be that hackers mainly operate anonymously and are difficult to reach. This paper aims to shed light on this interesting and relevant research question. In order to find answers, we conducted a survey among the members of the German Chaos Computer Club. As one of the world's largest organisations for information security and hacking, the club provides a good basis for our study. We examine the question of which programming languages are used by hackers as well as the importance of the programming language for their work. The paper offers first insights into the topic and can provide a starting point for further research.
\end{abstract}

\section{Introduction}
\label{sec:introduction}
In recent years, the Python programming language has gained the reputation to be popular among hackers \cite[p.~v]{Sarwar2021}. Since Python was one of the most widely used languages in 2021, this may not come as a surprise \cite{StackOverflow2021}. There are several online resources that support the hypothesis \cite{Simplilearn2022}\cite{Analyticsinsight2021}\cite{calltutors2020}\cite{ubuntupit2021}. Unfortunately, there are hardly any scientific studies available on the prevalence of programming languages in the hacking community. Our paper aims to help closing this research gap. The question we try to answer is: \textit{Which programming languages do hackers use?}

In order to address this research question, we conducted a survey among the members of the German Chaos Computer Club (CCC). As "Europe's largest association of hackers"\cite{CCC2022}, the club provides a good basis for our study. Before we explain our approach and discuss the results, the term \emph{hacker} needs to be clarified. Since there is no standard notion in the scientific literature, we define the term as follows:

\begin{definition}
A hacker is someone who uses his/her technical expertise to deal with computers with special regard to their security. We explicitly refer to the Chaos Computer Club's hacker ethics \cite{HackerEthics2022}. These form the core of the club's hacker definition and consequently that of our study.
\label{def:hacker}
\end{definition}

Some authors distinguish between "black-hat", "white-hat" and "grey-hat" hackers with respect to their ethics \cite[p.~20]{Marushat2019} \cite[pp.~12-13]{Sarwar2021}. By this definition, a white-hat hacker has no criminal intent, while black-hats use their computer knowledge for unlawful activities. Grey-hats are in between these two concepts. Raymond's guide on how to become a hacker uses the term "cracker" to describe malicious behaviour. According to Raymond, "hackers build things, crackers break them" \cite{Raymond2020}. Since we conducted our survey at the Chaos Computer Club, our research focuses on their hacker ethics, in line with Definition~\ref{def:hacker}. The club's hacker concept is not based on the black-hat/white-hat definition outlined above, but constitutes separate principles such as freedom of information and protection of private data \cite{HackerEthics2022}.

There are numerous  works on hacking in the (scientific) literature. Not all of them draw a line between different groups of hackers. Whether this distinction is important for the choice of programming language is another question. Programming languages are mainly discussed from a functional standpoint in literature. Raymond, for example, mentions Python, C, Perl, and Lisp as adequate choices \cite{Raymond2020}. Ericson provides code snippets \mbox{in C}\cite{Ericson2008}. Simpson and Antill describe C as "one of the most popular programming languages for security professionals and hackers" \cite[p.~196]{SimpsonAntill_2017}. The authors also write that hackers "use Perl to create automated exploits and malicious bots" \cite[p.~178]{SimpsonAntill_2017}. Clark explains hacking techniques with shell scripts under Linux and Windows as well as programmatic approaches with Python and other languages \cite{Clark2014}. Python is covered by a variety of further hacking books \cite{O_Connor_2012}, \cite{Seitz2015}, \cite{Sinha_2017}. Since software vulnerabilities play an import role for security, Turner analysed C, Java, C++, Objective-C, C\#, PHP, Visual Basic, Python, Perl and Ruby in this regard \cite{Turner_2014}.

As mentioned above, it is hard to find literature on the prevalence of programming languages in the hacking community. One reason might be that hackers often operate anonymously and are difficult to reach. Samtani et al. sidestepped this problem by exploring hacker assets in underground forums \cite{Samtani2015}. The authors used machine learning algorithms to classify whether code postings were written in Java, Python, C/C++, HTML, PHP, Delphi, ASP, SQL, Ruby, or Perl. This way, they could show that XSS attacks were primarily implemented in Perl, password cracks and keyloggers in Java, and finally banking vulnerabilities and Microsoft exploits in SQL \cite[p.~35]{Samtani2015}. Our paper follows an alternative approach to shed light on this topic.

\section{Approach}
\label{sec:approach}
In May 2021 we conducted a cross-sectional survey \cite{mcmillan1996educational} at the Chaos Computer Club. For this purpose, we have sent a link to an online questionnaire to the local and regional affiliates of the CCC (so-called Erfa-Kreise) \cite{ErfaKreise2022}. The design of the questionnaire was inspired by the 2019 Stack Overflow Developer Survey \cite{StackOverflow2019}. Other sources of inspiration were the Kaggle Data Science Survey \cite{Kaggle2020} and Raymond's guide on how to become a hacker \cite{Raymond2020}. Our questions focused on programming languages as well as related topics such as operating systems and development environments. For a better interpretation of the results, we also asked the participants how important they consider the choice of programming language to be for hacking.

In our questionnaire, we used Likert scales as well as multiple choices and \mbox{drop-downs}. Control questions were included to ensure proper answering \cite{raithel2008quantitative}. The survey instrument was pre-tested on 3 selected participants. Based on the pre-test we slightly revised some survey items, especially enclosed answer options and added examples and instructions.

We opened the questionnaire on 1 May and closed it on 30 May 2021. In total, we received 43 responses. As not all questions were mandatory, the number of responses for a given question may be lower than the total number of participants. It is clear that the results do not allow for a representative conclusion on our research question. Nevertheless, the study offers first insights into the topic and can serve as a starting point for further, broader research.

\newpage{}

\section{Discussion of Results}
\label{sec:results}
In the following, we discuss the most important results of our survey. For a better overview, we have divided the responses into several subsections.

\subsection{Experience}
Our first questions focused on the experience of the participants. Table~\ref{table:experience_general} shows that the majority of respondents had a general programming experience of several years. A slightly different picture emerged when we asked about the specific hacking experience as presented in Table~\ref{table:experience_hacking}.


\begin{table}[H]
\caption{How many years of programming experience do you have in general (hacking and non-hacking)?}
\centering
\setlength{\tabcolsep}{1,2em}\setlength\extrarowheight{2pt}
\begin{tabular}{|L{6cm}|R{3cm}|R{3cm}|}
\hline
\rowcolor{LightGray}
\thead{\textbf{Options}} &
\thead{\textbf{\% Percentages}} &
\thead{\textbf{\# Responses}}\\
\hline
20+ years & 41.86 \% & 18\\
\hline
10 - 20 years & 32.56 \% & 14\\
\hline
5 - 10 years & 18.60 \% & 8\\
\hline
3 - 5 years & 2.33 \% & 1\\
\hline
1 - 3 years & 4.65 \% & 2\\
\hline
$>$ 1 year & 0.00 \% & 0\\
\hline
I have never written code & 0.00 \% & 0\\
\hline
\cellcolor[gray]{0.8}Total & \cellcolor[gray]{0.8} & \cellcolor[gray]{0.8}43\\
\hline
\end{tabular}
\label{table:experience_general}
\end{table}


\begin{table}[H]
\caption{How many years of hacking experience do you have?}
\centering
\setlength{\tabcolsep}{1,2em}\setlength\extrarowheight{2pt}
\begin{tabular}{|L{6cm}|R{3cm}|R{3cm}|}
\hline
\rowcolor{LightGray}
\thead{\textbf{Options}} &
\thead{\textbf{\% Percentages}} &
\thead{\textbf{\# Responses}}\\
\hline
20+ years & 27.50 \% & 11\\
\hline
10 - 20 years & 27.50 \% & 11\\
\hline
5 - 10 years & 20.00 \% & 8\\
\hline
3 - 5 years & 15.00 \% & 6\\
\hline
1 - 3 years & 2.50 \% & 1\\
\hline
$>$ 1 year & 5.00 \% & 2\\
\hline
I have no hacking experience & 2.50 \% & 1\\
\hline
\cellcolor[gray]{0.8}Total & \cellcolor[gray]{0.8} & \cellcolor[gray]{0.8}40\\
\hline
\end{tabular}
\label{table:experience_hacking}
\end{table}

\subsection{Programming Language}
When asked about the programming languages used for hacking in the last year, participants named a variety of technologies. Table~\ref{table:languages_last_year} lists the responses in descending order. 


\begin{table}[H]
\caption{Which programming languages have you used for hacking in the last
year? (You can choose more than one answer option)}
\centering
\setlength{\tabcolsep}{1,2em}\setlength\extrarowheight{2pt}
\begin{tabular}{|L{6cm}|R{3cm}|R{3cm}|}
\hline
\rowcolor{LightGray}
\thead{\textbf{Options}} &
\thead{\textbf{\% Percentages}} &
\thead{\textbf{\# Responses}}\\
\hline
Bash/Shell/PowerShell & 72.50 \%	& 29\\
\hline
Python & 70.00 \% & 28\\
\hline
C & 32.50 \% & 13\\
\hline
JavaScript & 32.50 \% & 13\\
\hline
HTML/CSS & 30.00 \% & 12\\
\hline
C++ & 27.50 \% & 11\\
\hline
Go & 22.50 \% & 9\\
\hline
SQL & 22.50 \% & 9\\
\hline
Java & 20.00 \% & 8\\
\hline
Others & 20.00 \% & 8\\
\hline
Assembly & 17.50 \% & 7\\
\hline
C\# & 15.00 \% & 6\\
\hline
PHP & 15.00 \% & 6\\
\hline
Rust & 12.50 \% & 5\\
\hline
Ruby & 10.00 \% & 4\\
\hline
Perl & 7.50 \% & 3\\
\hline
TypeScript & 7.50 \% & 3\\
\hline
Kotlin & 5.00 \% & 2\\
\hline
Scala & 5.00 \% & 2\\
\hline
VB/VBA & 5.00 \% & 2\\
\hline
Lisp & 2.50 \% & 1\\
\hline
Swift & 2.50 \% & 1\\
\hline
Objective-C & 0.00 \% & 0\\
\hline
R & 0.00 \% & 0\\
\hline
\cellcolor[gray]{0.8}Total Respondents 40& \cellcolor[gray]{0.8} & \cellcolor[gray]{0.8}\\
\hline
\end{tabular}
\label{table:languages_last_year}
\end{table}

The options of programming languages provided in the questionnaire were based on a Stack Overflow survey \cite{StackOverflow2019}, Raymond's hacking guide \cite{Raymond2020} and feedback from our pre-testers. Whether technologies such as HTML, Bash or SQL can be called programming languages is of course debatable. We have included them in the list anyway to avoid possible gaps in the study.

According to Table~\ref{table:languages_last_year}, shell scripts (e.g. Bash) and Python were used most frequently. It also appears that the C language family (C, C++, C\#, and Objective-C) is still common. With regard to Java, the survey supports Raymond's argument that this language is not the first choice for hackers \cite{Raymond2020}. Table~\ref{table:languages_preference} shows that the language preference of the majority of participants has changed over time. Compared to programming languages used more than a year before (see Table~\ref{table:languages_more_years} in the appendix), a shift towards shell scripts and Python can be noted. Python's rise in popularity has been observed by other developer surveys too \cite{StackOverflow2019} \cite{StackOverflow2021}.


\begin{table}[H]
\caption{Has your programming language preference changed over time?}
\centering
\setlength{\tabcolsep}{1,2em}\setlength\extrarowheight{2pt}
\begin{tabular}{|L{6cm}|R{3cm}|R{3cm}|}
\hline
\rowcolor{LightGray}
\thead{\textbf{Options}} &
\thead{\textbf{\% Percentages}} &
\thead{\textbf{\# Responses}}\\
\hline
yes, my preference has changed & 77.50 \% & 31\\
\hline
no, I always used the same programming languages & 22.5 \% & 9\\
\hline
\cellcolor[gray]{0.8}Total & \cellcolor[gray]{0.8} & \cellcolor[gray]{0.8}40\\
\hline
\end{tabular}
\label{table:languages_preference}
\end{table}


\begin{table}[H]
\caption{How strongly do you agree with the following statement: "The choice of
the programming language is important for hacking."}
\centering
\setlength{\tabcolsep}{1,2em}\setlength\extrarowheight{2pt}
\begin{tabular}{|L{6cm}|R{3cm}|R{3cm}|}
\hline
\rowcolor{LightGray}
\thead{\textbf{Options}} &
\thead{\textbf{\% Percentages}} &
\thead{\textbf{\# Responses}}\\
\hline
Strongly Agree & 5.0 \% & 2\\
\hline
Agree & 20.0 \% & 8\\
\hline
Neither / Nor Agree & 32.50 \% & 13\\
\hline
Disagree & 20.00 \% & 8\\
\hline
Strongly Disagree & 22.50 \% & 9\\
\hline
\cellcolor[gray]{0.8}Total & \cellcolor[gray]{0.8} & \cellcolor[gray]{0.8}40\\
\hline
\end{tabular}
\label{table:languages_importance}
\end{table}

An interesting result appears in Table~\ref{table:languages_importance}. Many respondents (75\%) did not agree that the choice of the programming language is important for hacking. Only 25\% of participants agreed or strongly agreed with this statement. The prevalence of Python for hacking might therefore simply reflect the general increase in its use in recent years. Consequently, one could expect that the language preference of hackers will continue to change in future as technology evolves.

\subsection{Ecosystem}
We also asked the participants which operating systems (OS) they used for hacking in the last year. Table~\ref{table:operating_systems} shows that the majority of respondents chose a Linux-based variant. This is not surprising, since Kali Linux even provides a specific distribution for security and penetration testing \cite{Kali2022}. When asked about the integrated development environments (IDEs), a variety of tools were selected. As evident from Table~\ref{table:ides}, Vim and Visual Studio Code were in the top five (see Table~\ref{table:ides_complete} in the appendix for a full list).


\begin{table}[H]
\caption{Which operating systems have you used for hacking in the last year? (You can choose more than one answer option)}
\centering
\setlength{\tabcolsep}{1,2em}\setlength\extrarowheight{2pt}
\begin{tabular}{|L{6cm}|R{3cm}|R{3cm}|}
\hline
\rowcolor{LightGray}
\thead{\textbf{Options}} &
\thead{\textbf{\% Percentages}} &
\thead{\textbf{\# Responses}}\\
\hline
Linux-based & 95.00 \% & 38\\
\hline
Windows & 40.00 \% & 16\\
\hline
MacOS & 32.50 \% & 13\\
\hline
BSD & 17.50 \% & 7\\
\hline
Others & 5.00 \% & 2\\
\hline
\cellcolor[gray]{0.8}Total Respondents 40& \cellcolor[gray]{0.8} & \cellcolor[gray]{0.8}\\
\hline
\end{tabular}
\label{table:operating_systems}
\end{table}


\begin{table}[H]
\caption{Which integrated development environments (IDEs) have you used for hacking in the last year?(You can choose more than one answer option)}
\centering
\setlength{\tabcolsep}{1,2em}\setlength\extrarowheight{2pt}
\begin{tabular}{|L{6cm}|R{3cm}|R{3cm}|}
\hline
\rowcolor{LightGray}
\thead{\textbf{Options}} &
\thead{\textbf{\% Percentages}} &
\thead{\textbf{\# Responses}}\\
\hline
Vim & 60.00 \% & 24\\
\hline
Visual Studio Code & 50.00 \% & 20\\
\hline
Others & 22.50 \% & 9\\
\hline
IntelliJ & 17.50 \%  & 7\\
\hline
Visual Studio & 17.50 \% & 7\\
\hline
... & ... & ...\\
\hline
\cellcolor[gray]{0.8}Total Respondents 40& \cellcolor[gray]{0.8} & \cellcolor[gray]{0.8}\\
\hline
\end{tabular}
\label{table:ides}
\end{table}

\subsection{Demographics} Finally, we asked the participants about their gender identity and age. Table~\ref{table:gender} shows that most respondents identified themselves as male. Their age varied, with the majority between 25 and 44 years old, as revealed by Table~\ref{table:age}.


\begin{table}[H]
\caption{What is your gender identity?}
\centering
\setlength{\tabcolsep}{1,2em}\setlength\extrarowheight{2pt}
\begin{tabular}{|L{6cm}|R{3cm}|R{3cm}|}
\hline
\rowcolor{LightGray}
\thead{\textbf{Options}} &
\thead{\textbf{\% Percentages}} &
\thead{\textbf{\# Responses}}\\
\hline
woman & 2.50 \% & 1\\
\hline
man & 60.00 \% & 24\\
\hline
nonbinary & 12.50 \% & 5\\
\hline
prefer not to say & 20.00 \% & 8\\
\hline
prefer to self-describe & 5.00 \% & 2\\
\hline
\cellcolor[gray]{0.8}Total& \cellcolor[gray]{0.8} & \cellcolor[gray]{0.8}40\\
\hline
\end{tabular}
\label{table:gender}
\end{table}


\begin{table}[H]
\caption{What is your age?}
\centering
\setlength{\tabcolsep}{1,2em}\setlength\extrarowheight{2pt}
\begin{tabular}{|L{6cm}|R{3cm}|R{3cm}|}
\hline
\rowcolor{LightGray}
\thead{\textbf{Options}} &
\thead{\textbf{\% Percentages}} &
\thead{\textbf{\# Responses}}\\
\hline
0 - 17 & 2.50 \% & 1\\
\hline
18 - 21 & 2.50 \% & 1\\
\hline
22 - 24 & 5.00 \% & 2\\
\hline
25 - 29 & 17.50 \% & 7\\
\hline
30 - 34 & 22.50 \% & 9\\
\hline
35 - 39 & 15.00 \% & 6\\
\hline
40 - 44 & 20.00 \% & 8\\
\hline
45 - 49 & 7.50 \% & 3\\
\hline
50 - 54 & 2.50 \% & 1\\
\hline
55 - 59 & 5.00 \% & 2\\
\hline
60 - 69 & 0.00 \% & 0\\
\hline
70+ & 0.00 \% & 0\\
\hline
\cellcolor[gray]{0.8}Total & \cellcolor[gray]{0.8} & \cellcolor[gray]{0.8}40\\
\hline
\end{tabular}
\label{table:age}
\end{table}

\pagebreak

\section{Conclusion}
\label{sec:conclusion}
The purpose of this paper was to shed light on the question of which programming languages are used by hackers. In order to achieve that goal, we conducted a survey at the German Chaos Computer Club in May 2021. Our results show that the members were using different programming languages at the time. Shell scripts and Python were chosen most frequently. It also appears that the C language family is still common. Another important finding is that the choice of programming language does not play a vital role for hackers. Their language preference has changed over time and will presumably continue to do so in the future.

The number of responses we received does not allow for a representative conclusion. Furthermore, the survey targeted only members of the CCC. The findings might therefore be biased both regionally and in favour of a specific group. Our results do, however, add to the extremely scarce literature on the subject. The approach could serve as a model for future surveys, possibly at international level.

\newpage
\appendix
\section{Additional Questions}

This section contains responses to our questionnaire that were referenced but not included in the main text.


\begin{table}[H]
\caption{If your programming language preference has changed, which
programming languages have you used for hacking more than a year ago?
(You can choose more than one answer option)}
\centering
\setlength{\tabcolsep}{1,2em}\setlength\extrarowheight{2pt}
\begin{tabular}{|L{6cm}|R{3cm}|R{3cm}|}
\hline
\rowcolor{LightGray}
\thead{\textbf{Options}} &
\thead{\textbf{\% Percentages}} &
\thead{\textbf{\# Responses}}\\
\hline
Bash/Shell/PowerShell & 47.06 \% & 16\\
\hline
Python & 41.18 \% & 14\\
\hline
C & 38.24 \% & 13\\
\hline
PHP & 32.35 \% & 11\\
\hline
C++ & 26.47 \% & 9\\
\hline
JavaScript & 26.47 \% & 9\\
\hline
Assembly & 23.53 \% & 8\\
\hline
Java & 23.53 \% & 8\\
\hline
Others & 20.59 \% & 7\\
\hline
HTML/CSS & 17.65 \% & 6\\
\hline
Perl & 14.71 \% & 5\\
\hline
SQL & 8.82 \% & 3\\
\hline
VB/VBA & 8.82 \% & 3\\
\hline
C\# & 5.88 \% & 2\\
\hline
Rust & 5.88 \%  & 2\\
\hline
R & 2.94 \% & 1\\
\hline
Ruby & 2.94 \% & 1\\
\hline
Go & 0.00 \% & 0\\
\hline
Kotlin & 0.00 \% & 0\\
\hline
Lisp & 0.00 \% & 0\\
\hline
Objective-C & 0.00 \% & 0\\
\hline
Scala & 0.00 \% & 0\\
\hline
Swift & 0.00 \% & 0\\
\hline
TypeScript & 0.00 \% & 0\\
\hline
\cellcolor[gray]{0.8}Total Respondents 34& \cellcolor[gray]{0.8} & \cellcolor[gray]{0.8}\\
\hline
\end{tabular}
\label{table:languages_more_years}
\end{table}

\begin{table}[H]
\caption{Which integrated development environments (IDEs) have you used for hacking in the last year? (You can choose more than one answer option)}
\centering
\setlength{\tabcolsep}{1,2em}\setlength\extrarowheight{2pt}
\begin{tabular}{|L{6cm}|R{3cm}|R{3cm}|}
\hline
\rowcolor{LightGray}
\thead{\textbf{Options}} &
\thead{\textbf{\% Percentages}} &
\thead{\textbf{\# Responses}}\\
\hline
Vim & 60.00 \% & 24\\
\hline
Visual Studio Code & 50.00 \% & 20\\
\hline
Others & 22.50 \% & 9\\
\hline
IntelliJ & 17.50 \%  & 7\\
\hline
Visual Studio & 17.50 \% & 7\\
\hline
Android Studio & 15.00 \% & 6\\
\hline
Eclipse & 15.00 \% & 6\\
\hline
Nano & 15.00 \% & 6\\
\hline
Notepad++ & 15.00 \% & 6\\
\hline
PyCharm & 12.50 \% & 5\\
\hline
Sublime Text & 12.50 \% & 5\\
\hline
Atom & 7.50 \% & 3\\
\hline
IPython / Jupyter & 7.50 \% & 3\\
\hline
NetBeans & 5.00 \% & 2\\
\hline
PHPStorm & 5.00 \% & 2\\
\hline
RubyMine & 5.00 \% & 2\\
\hline
Xcode & 5.00 \% & 2\\
\hline
Coda & 2.50 \% & 1\\
\hline
Emacs & 2.50 \% & 1\\
\hline
TextMate & 2.50 \% & 1\\
\hline
Komodo & 0.00 \% & 0\\
\hline
RStudio & 0.00 \% & 0\\
\hline
\cellcolor[gray]{0.8}Total Respondents 40& \cellcolor[gray]{0.8} & \cellcolor[gray]{0.8}\\
\hline
\end{tabular}
\label{table:ides_complete}
\end{table}

\section{Glossary}

\label{appx:glossary}
\begin{table}[H]
\centering
\setlength{\tabcolsep}{1,2em}\setlength\extrarowheight{3pt}
\begin{tabular}{|l|p{9cm}|}
\hline
\thead{\textbf{Term}} &
\thead{\textbf{Definition}}\\
\hline
CCC &
Chaos Computer Club\\
\hline
Cracker &
Individual using computer knowledge with malicious intent\\
\hline
Erfa-Kreise &
Local and regional affiliates of the Chaos Computer Club\\
\hline
Hacker &
See Definition~\ref{def:hacker}\\
\hline
Keylogger &
Software for keystroke recording\\
\hline
XSS &
Cross site scripting\\
\hline
\end{tabular}
\end{table}

\newpage
\vspace{1em}
\bibliographystyle{unsrt}
\bibliography{main}

\end{document}